# Large anomalous unidirectional magnetoresistance in a single ferromagnetic layer

Kaihua Lou[1,2], Qianwen Zhao[1,2], Baiqing Jiang[1,2], and Chong Bi[1,2*]

[1]The Key Laboratory of Microelectronics Device & Integrated Technology, Institute of Microelectronics Chinese Academy of Sciences, Beijing 100029, China

[2]University of Chinese Academy of Sciences, Beijing 100049, China

* bichong@ime.ac.cn

**Abstract**

Unidirectional magnetoresistance (UMR) in a ferromagnetic bilayer due to the spin Hall effects (SHEs) provides a facile means of probing in-plane magnetization to avoid complex magnetic tunnel junctions. However, the UMR signal is very weak and usually requires a lock-in amplifier for detection even in the bilayer involving Ta or Pt with a large spin Hall angle (SHA). Here we report a type of UMR, termed as the anomalous UMR (AUMR), in a single CoFeB layer without any adjacent SHE layers, where the UMR signal is about 10 times larger than that in Ta/CoFeB structures and can be detected by using conventional dc multimeters in the absence of lock-in amplifiers. We further demonstrate that the extracted AUMR by excluding thermal contributions shows reversal signs for the CoFeB and NiFe single layers with opposite SHAs, indicating that the AUMR may originate from the self-generated spin accumulation interacting with magnetization through the giant magnetoresistance-like mechanism. These results suggest that the AUMR contributes UMR signals larger than the interfacial spin Hall UMR in the CoFeB-involved systems, providing a convenient and reliable approach to detect in-plane magnetization for the two-terminal spintronic devices.

# I. Introduction

Electrical detection of magnetization is a central theme in spintronics [1]. In addition to the giant (GMR) [2,3] or tunnel magnetoresistance (TMR) [4] effects in complex magnetic tunnel junction (MTJ) structures, the magnetization can also be detected easily by measuring the anomalous Hall effect [5] (AHE) in a perpendicularly magnetized ferromagnet. However, for an in-plane magnetized ferromagnet, there is no such facile approach to detect magnetization except by employing MTJs [6]. Anisotropic magnetoresistance (AMR, $\propto cos^2\alpha$, with $\alpha$ being the angle between sensing current and magnetization direction) and the planar Hall effects (PHE, $\propto \sin 2\alpha$) result in the same resistance for a 180° magnetization reversal [7], and thus directly probing in-plane magnetization via both effects is impossible. Usually, the MTJ-based magnetization detection requires extremely strict sputtering conditions along with the following multistep nanofabrication processes [6,8,9]. Recent studies have suggested that an additional spin current-induced unidirectional spin Hall magnetoresistance (USMR) emerging in ferromagnet/heavy-metal (FM/HM) bilayers due to the spin Hall effects (SHEs) [10,11] may provide an alternative means to electrical readout of the in-plane magnetization by using a two-terminal geometry [12–15]. USMR arises from the modulation of interfacial resistance due to the SHE-induced spin accumulation at the FM/HM interface through the spin-dependent electron scattering or electron-magnon scattering [11,16,17]. In this scheme, an adjacent HM or other materials showing large spin Hall angle (SHA) as the spin polarizer is required to obtain the observable USMR [10,11,13,18,19]. From the point of view of experiments, USMR is the second order of nonlinear resistance, which is usually detected by using highly sensitive lock-in techniques even for the samples with strong SHE [10,11,18].

In this work, we report a type of unidirectional magnetoresistance (UMR) in a single ferromagnet without any adjacent SHE materials or the ferromagnetic reference layers, termed as the anomalous unidirectional magnetoresistance (AUMR). By employing a single CoFeB layer, we demonstrate that the AUMR signal is even larger than USMR signal in the CoFeB/HM bilayers and can be measured reliably by using conventional dc multimeters without lock-in amplifiers. We further extract AUMR from thermal signals and find that it is opposite for the CoFeB and NiFe single layers with opposite SHAs [20] and spin-orbit torques (SOTs) [21]. These results indicate that the large AUMR in a single CoFeB layer originates from the self-generated spin accumulation due to spin-orbit coupling and can be facilely used to investigate current-induced in-plane magnetization switching and SOTs generated in other materials that are not suitable for sputtering MTJs, such as 2D materials [22,23], topological materials [24–27], and axion insulators [28].

## II. Methods

### A. Sample fabrication

CoFeB is a widely used ferromagnetic material in TMR devices, for example, as the ferromagnetic electrodes in MgO-based MTJs to get high TMR [29,30]. Since the discovery of USMR [10], the CoFeB has also been used to evaluate spin-orbit effects by measuring USMR [13,18], but simply assuming that the USMR signal from the CoFeB itself arises from the thermal effects. To convincingly demonstrate AUMR in the CoFeB single layer, in addition to the conventional measures for excluding possible artificial thermal effects, it is desirable to observe AUMR in other ferromagnetic materials with opposite AUMR signs while keeping other possible artificial effects the same. Therefore, another ferromagnetic material, NiFe with opposite SHAs [20] and SOTs [21], is chosen as the control material to observe the opposite AUMR. Other control samples including Co (10), Sub/Pt (6)/CoFeB (10), Sub/CoFeB (10)/Pt (6), Sub/Ta (8)/CoFeB (10), and Sub/CoFeB (10)/Ta (8) where the numbers in parentheses represent thicknesses ($t$) in the unit of nm are also deposited and tested for clarifying the AUMR effects. The CoFeB (5 - 30) layers as well as control samples are deposited on the Si/SiO$_2$ (300) substrates. The capping layers are 10 nm SiO$_2$ sputtered from a SiO$_2$ target for all samples to prevent further oxidation. All layers are deposited by magnetron sputtering with a base pressure of $4.5 \times 10^{-4}$ Pa and the purity of all sputtering targets is higher than 99.95%. The as-deposited films are then patterned into micron-sized Hall-bar structures with nominal length $l$ = 50 - 100 $\mu$m and width $w$ = 2 - 10 $\mu$m by using standard photolithography and subsequent Ar-ion milling processes. The structure and element distribution of the single CoFeB layer are also investigated by high-resolution transmission electron microscopy (HRTEM) and scanning transmission electron microscopy (STEM) equipped with an energy dispersive X-ray spectroscopy (EDS). The cross-section HRTEM samples are prepared by using a focused-ion-beam system and then transferred into the microscope immediately to avoid possible oxidization in air.

### B. Electrical measurements

To measure the field-dependent AUMR, we first apply a positive current and then a negative current to measure the resistance as a function of the applied magnetic field (H). AUMR is then calculated by using $\Delta R = R(I^+) - R(I^-)$. For the angle-dependent measurements, the samples are mounted on a rotation sample holder controlled by a stepper motor. The applied field is 6 kOe and current is ±1.8 mA for all angle-dependent measurements. For the current-driven magnetization switching, we first apply a 1 ms current pulse ($I_p$) to switch the magnetization and then apply ±1.5 mA currents to measure AUMR after waiting for 3 s, similar to the measurement procedures for the current-driven perpendicular magnetization switching [31,32]. In all electrical measurements, the current is provided through a Keithley 6221 current source and the voltage is monitored by a Keithley 2000 multimeter. The magnetic field is generated through a homemade electromagnet powered by a Kepco bipolar power supply.

## III. Results and discussion

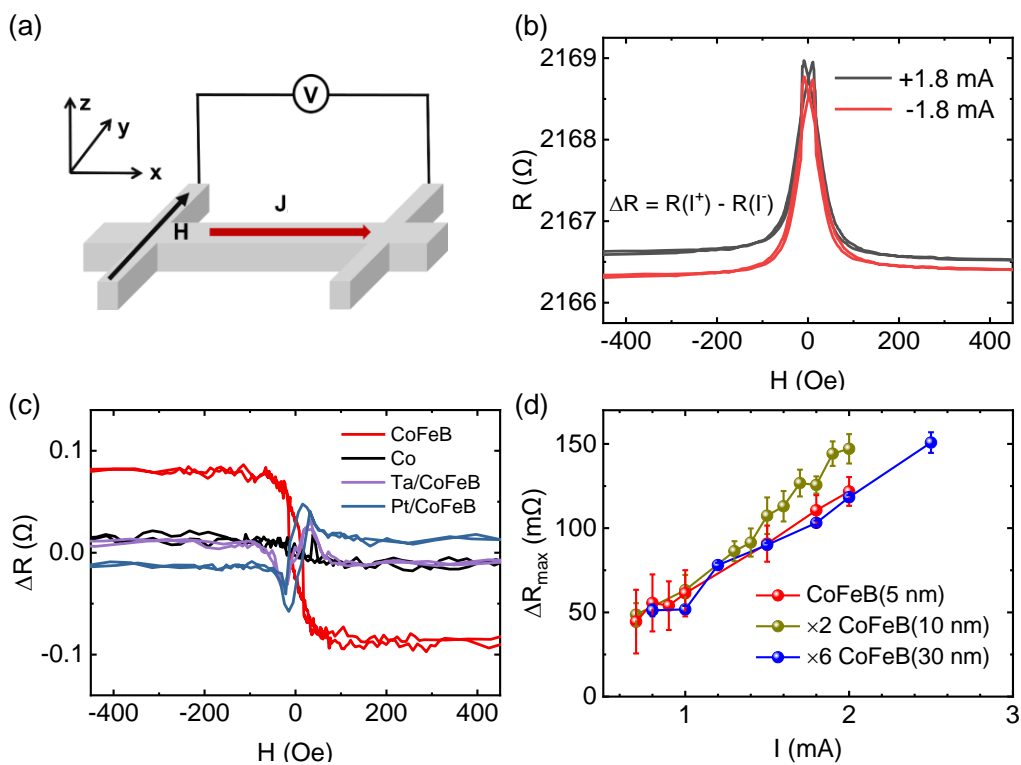

FIG. 1. Experimental configurations and AUMR measurement results. (a) Schematics of the measurement configurations. (b) Typical field-dependent resistance of the 10 nm CoFeB layers measured by using I = ±1.8 mA. The resistance difference ΔR is defined as $R(I^+) - R(I^-)$. The little entire shift between +1.8 mA and -1.8 mA curves arises from the thermal effect due to Joule

heating, which is not exactly the same for the positive and negative applied current as demonstrated below. (c) Recorded ΔR as a function of H for CoFeB(10 nm), Co (10 nm), Ta(8 nm)/CoFeB(10 nm), and Pt(6 nm)/CoFeB(10 nm) with $l$ = 50 μm, $w$ = 5 μm by using I = ±1.8 mA. All curves are centered to ΔR = 0 Ω. (d) Current-dependent $\Delta R_{max}$ for the 5 nm, 10 nm, and 30 nm CoFeB layers with $l$ = 50 μm, $w$ = 10 μm. The signals for the 10 nm and 30 nm samples are magnified by 2 and 6 times, respectively.

## A. AUMR measurements in CoFeB and control samples

As illustrated in Fig. 1(a), the applied current (I) is along the *x*-direction and H is applied along the *y* direction. We first record the resistance (R) as a function of H, which shows typical AMR characteristics as shown in Fig. 1(b), and then evaluate the AUMR effect by using $\Delta R = R(I^+) - R(I^-)$. Generally, according to the AMR theory [7], R should not depend on the applied current direction and $R(I^+) = R(I^-)$. However, besides a little entire shift due to unequal Joule heating with $I^+$ and $I^+$, Fig. 1(b) shows that the $R(I^+)$ and $R(I^-)$ curves are significantly different in the positive and negative field ranges for the CoFeB layer. For I = +1.8 mA, the resistance in the negative field range ($R|_{H<-100\,Oe}$) is larger than that in the positive field range ($R|_{H>100\,Oe}$), while for I = -1.8 mA, $R|_{H<-100\,Oe} < R|_{H>100\,Oe}$. The H-dependent resistance change for I = ±1.8 mA cannot be attributed to the resistance increase due to Joule heating, which should not relate to H and can only induce the entire shift of resistance curves. To reflect the H-dependent resistance change directly, ΔR is calculated as shown in Fig. 1(c), where ΔR strongly depends on the magnetized direction and saturates to a positive and a negative value when |H| > 100 Oe, indicating a strong in-plane magnetization dependence. To compare to the reported USMR effects [10], Fig. 1(c) also presents ΔR measurements for the Co single layer, Pt/CoFeB and Ta/CoFeB bilayers. It has been shown that the maximum of ΔR, $\Delta R_{max}$, defined as $\Delta R|_{H<-100\,Oe} - \Delta R|_{H>100\,Oe}$, for the single CoFeB layer is about 5 and 10 times larger than that for the Pt/CoFeB and Ta/CoFeB bilayers, respectively. Even we assume that the current distributes in the Pt/CoFeB and Ta/CoFeB bilayers uniformity, ΔR of the single CoFeB layer is still much larger at the same current density. The two ΔR peaks around ±25 Oe for the Pt/CoFeB and Ta/CoFeB bilayers are due to the magnon scattering at the FM/HM interfaces as reported before [11]. This is also verified by the ΔR curves of the Co and CoFeB single layers where no such peaks appear because of the absence of FM/HM interfaces.

Figure 1(d) shows $\Delta R_{max}$ as a function of I for the single CoFeB layer with different thicknesses. Surprisingly, in addition to the linear current dependence, $\Delta R_{max}$ is reversely proportional to the thickness at the same current. For example, $\Delta R_{max}$ of the 5 nm CoFeB layer is about 2 and 6 times larger than that of the 10 nm and 30 nm CoFeB, respectively. These results indicate that $\Delta R_{max}$ is determined by the applied current density ($J$) regardless of the CoFeB thickness, which strongly conflicts with a thermal origin due to the Joule heating such as the anomalous Nernst effect [33,34] (ANE) or the spin Seebeck effect (SEE) [35]. This is because the temperature increase ($\Delta T$) induced by Joule heating should not be exactly the same for different CoFeB thicknesses at the same current density, especially by considering different resistivities and thermal conductivities between the CoFeB and surrounding layers, as confirmed below.

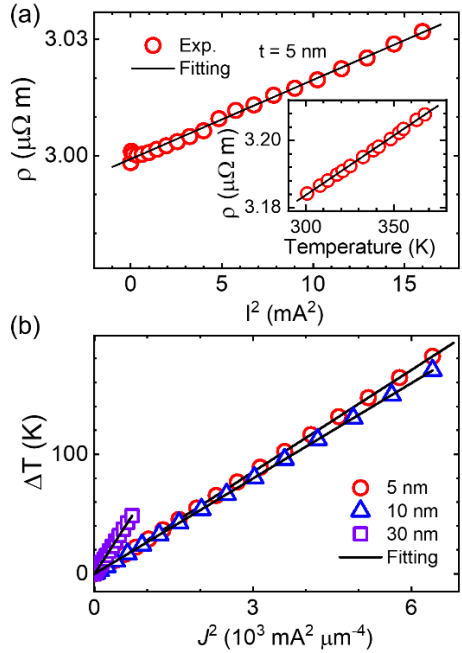

FIG. 2. Calibration of temperature increase induced by Joule heating. (a) The resistivity of 5 nm CoFeB as a function of $I^2$. The resistance was recorded after waiting 3 s for each current step. Inset shows the corresponding temperature-dependent resistivity. (b) The calibrated temperature increase as a function of applied current density for the CoFeB layers with different thicknesses. The solid lines are linear fitting results.

**B. Calibration of thermal effects induced by Joule heating**

To evaluate $\Delta T$ induced by Joule heating, we first calibrate the temperature coefficient of resistivity for different CoFeB thicknesses and then use the calibrated temperature coefficient to evaluate temperature increase by measuring the resistivity as a function of $I^2$. The corresponding experimental results are shown in Fig. 2(a), from which $\Delta T$ is calculated (Fig. 2(b)). The measured $\Delta T$ shows a linear dependence on $J^2$, consistent with the Joule heating mechanism where $\Delta T \propto J^2 R$. As shown in Fig. 2(b), when t = 5 nm and 10 nm, the slopes of ΔT versus $J^2$ are almost the same but significantly smaller than that for $t$ = 30 nm. This may be understood by considering a similar heat dissipation capacity for very thin films ($t$ = 5 nm and 10 nm) through the substrates and capping layers, which becomes limited for thick films ($t$ = 30 nm) under the same heating current density. These results directly confirm that ΔT varies with CoFeB thickness at the same $J$ and demonstrate that the thermal effects cannot be the only origin for the measured $\Delta R_{max}$.

To quantitatively determine the thermal contribution to ΔR ($\Delta R^{Thermal}$), we have further measured the angle dependences of ΔR and the Hall resistance difference, $\Delta R_H = R_H(I^+) - R_H(I^-)$, as shown in Fig. 3(a) and 3(b). It has been approved that $\Delta R_H$ can be used to extract thermal contribution according to the $l/w$ ratio [36] since $\Delta R_H$ can only be induced by the pure thermal effects under an $x$-directional field. As shown in Fig. 3(a) and 3(b), R and $R_H$ present a sin $\alpha$ and cos $2\alpha$ dependence, respectively, which is the typical AMR and PHE characteristics. For the 10 nm CoFeB layer, the calculated $l/w$ ratio is about 6.67 from the variation of R and $R_H$ with the nominal $l$ = 50 μm and $w$ = 5 μm. In contrast, ΔR and $\Delta R_H$ show a sin $\alpha$ and cos $\alpha$ dependence, as expected from the UMR and thermal effects, respectively. Combining with the calculated $l/w$ ratio, $\Delta R^{Thermal}$ can be estimated through $\Delta R_{max}^{Thermal} = \Delta R_H^{max} l/w$, which is 97.38 mΩ, about 55.64% of the fitted $\Delta R_{max}$ (175.01 mΩ) for the 10 nm CoFeB layer. Here, $\Delta R_{max}$ and $\Delta R_H^{max}$ are the fitted peak-to-peak values of ΔR and $\Delta R_H$ by using the sin $\alpha$ and cos $\alpha$ functions in Fig. 3(a) and 3(b), respectively. It should be noted that the error of this method for quantifying thermal contributions is less than 10% in a ferromagnetic insulator/Pt bilayer with only pure thermal effects where no any other UMR contributes [37] because of the insulating ferromagnet. These results further confirm AUMR contribution to the measured ΔR ($\Delta R^{AUMR}$). The solid lines in Fig. 3(a) are fitting results by using $\Delta R = \Delta R^{AUMR} + \Delta R^{Thermal}$, and the fitted peak-to-peak value of $\Delta R^{AUMR}$ is about 77.63 mΩ. The AUMR ratio, $\Delta R^{AUMR}/R$, is about 3.68 × 10⁻⁵, with the same order as USMR in FM/HM system involving strong SHEs [10,11]. Table 1 summarizes the extracted $\Delta R^{AUMR}$ for different $l/w$ ratios, in which $\Delta R_{max}^{Thermal}$ is much less than $\Delta R_{max}$ for all CoFeB samples, indicating a positive $\Delta R^{AUMR}$. In addition, for the same thickness, the calculated AUMR ratio with the same current density does not depend on the $l/w$ ratio. For the CoFeB layer with $t$ = 5 nm, the calculated AUMR ratios are 4.47 × 10⁻⁵, 3.64 × 10⁻⁵, and 3.97 × 10⁻⁵ per current density of 10⁷ A cm⁻² for the three different $l/w$ ratios (8.46, 4.56, and 8.95, respectively). If we choose the medium value of $4.05 \times 10^{-5}$ per $10^7$ A cm$^{-2}$, the variation of measured AUMR ratios is within 10%. The independence of AUMR ratio on the *l/w* ratio is a typical magnetoresistance characteristic. For the NiFe samples with $t = 2.5$ nm, the calculated AUMR ratios are $-2.72 \times 10^{-5}$ and $-4.19 \times 10^{-5}$ per $10^7$ A cm$^{-2}$ with the *l/w* ratio of 9.15 and 4.58, respectively. The large variation of AUMR ratio for the NiFe sample may be because the UMR signals are much weaker compared to the CoFeB samples as shown below, in which the larger measurement errors are induced. The thickness dependence of AUMR ratio and reduction of AUMR after annealing, both of which relate to the bulk inversion symmetry breaking, as well as the negative AUMR sign for the NiFe layer will be discussed later.

The ΔR results for the Sub/Pt/CoFeB, Sub/CoFeB/Pt, Sub/Ta/CoFeB, and Sub/CoFeB/Ta have also been measured to qualitatively evaluate AUMR contribution from the USMR results in the CoFeB-involved bilayers. When the HM layer is on the top or at the bottom of the CoFeB layer, the SHE-induced USMR is opposite while the AUMR as well as $\Delta R^{Thermal}$ from CoFeB layer keeps the same. As shown in Fig. 3(c), for the Pt-CoFeB bilayers, ΔR in both the large and small field ranges are opposite with a top and bottom Pt layer, indicating that the SHE-induced USMR dominates ΔR measurement results. The weaker ΔR of the Sub/Pt/CoFeB sample can be understood that the sign of USMR is opposite to that of AUMR as well as the thermal effects. While for the Ta-CoFeB bilayers, $\Delta R|_{|H| > 100\text{ Oe}}$ always shows the same sign as that of the single CoFeB layer (Fig. 1c) whether the Ta layer is on the top or at the bottom of the CoFeB layer, indicating that AUMR and $\Delta R^{Thermal}$ of the CoFeB layer dominate the measured ΔR. This sharply contrasts to the Ta/Co bilayers where the USMR still dominates ΔR [10] because of the much weaker AUMR and $\Delta R^{Thermal}$ for the Co single layer compared to the CoFeB layer as shown in Fig. 1(c). It should be noted that $\Delta R|_{|H| < 100\text{ Oe}}$ due to SHE-modulated magnon scattering are always opposite for the Sub/HM/CoFeB and Sub/CoFeB/HM (HM: Ta or Pt) samples.

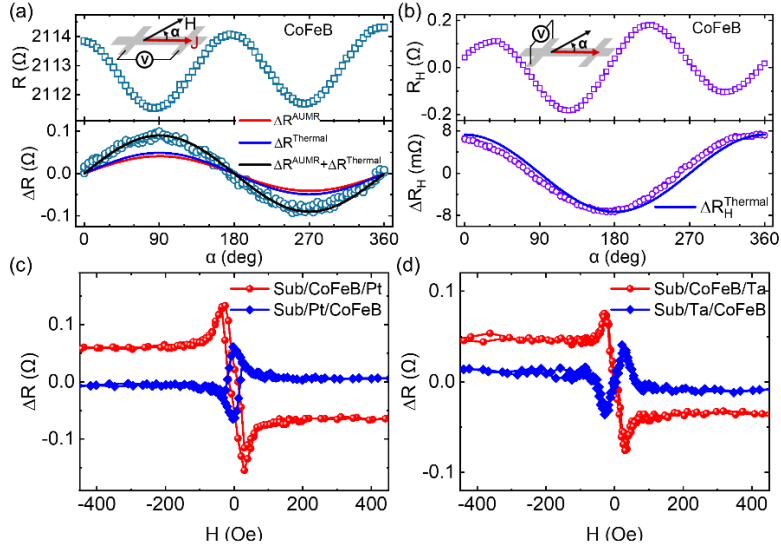

FIG. 3. Extraction of AUMR from thermal effects. (a), (b) Angle dependence of longitudinal (a) and transverse resistance (b) measurement results for the 10 nm CoFeB single layer with $l = 50$ μm, $w = 5$ μm. The applied field is 6 kOe and the applied current is ±1.8 mA. Insets show the measurement configurations schematically. The solid lines are the calculated and fitting results arising from thermal effects and AUMR. (c), (d) Field-dependent AUMR measurement results for the Sub/CoFeB/Pt, Sub/Pt/CoFeB (c), Sub/CoFeB/Ta, and Sub/Ta/CoFeB (d) structures.

TABLE 1. Summary of AUMR measurement results for different ferromagnetic materials with various nominal $l/w$ ratios. The applied current is ± 1.8 mA. CoFeB* indicates that the sample has been annealed at 200 °C for 10 minutes under $N_2$ atmosphere.

| FM | $t$ (nm) | $l$ (μm) | $w$ (μm) | Calculated $l/w$ | $\Delta R_H^{max}$ (mΩ) | $\Delta R_{max}$ (mΩ) | $\Delta R_{max}^{Thermal}/\Delta R_{max}$ (%) | AUMR (mΩ) | AUMR/R ($10^{-5}$) |
|---|---|---|---|---|---|---|---|---|---|
| CoFeB | 10 | 50 | 5 | 6.67 | 14.60 | 175.01 | 55.64 | 77.63 | 3.68 |
| CoFeB | 5 | 50 | 5 | 8.46 | 29.48 | 384.88 | 64.82 | 135.41 | 3.22 |
| CoFeB | 5 | 50 | 10 | 4.56 | 17.88 | 108.74 | 75.02 | 27.16 | 1.31 |
| CoFeB* | 5 | 50 | 10 | 4.65 | 14.77 | 78.57 | 87.39 | 9.91 | 0.53 |
| CoFeB | 5 | 100 | 10 | 8.95 | 15.40 | 197.38 | 69.82 | 59.57 | 1.43 |
| NiFe | 10 | 50 | 5 | 8.72 | 2.66 | 10.78 | 215.17 | -12.42 | -1.95 |
| NiFe | 2.5 | 50 | 5 | 9.15 | 35.98 | 148.76 | 221.25 | -180.37 | -3.91 |
| NiFe | 2.5 | 50 | 10 | 4.58 | 27.30 | 54.35 | 230.17 | -70.75 | -3.02 |

## C. Mechanisms and applications

Recently, it has been reported both theoretically [38] and experimentally that spin accumulation can be established even in a ferromagnet like non-magnetic SOT layers due to the counterpart of AHE or AMR [21,39–43]. Similar to the GMR [2,3] and USMR [10,11], we suggest that the large AUMR arises from the AHE- and AMR-generated spin accumulation interacting with the magnetization of the single ferromagnet layer, as illustrated in Fig. 4(a). When the spin polarization of the established spin accumulation in the ferromagnet is parallel or antiparallel with the magnetization, a resistance change like the GMR and USMR effects will be induced. To verify this mechanism, we have also measured both ΔR and $\Delta R_H$ in a single NiFe layer. For the NiFe, it has been shown that both the SHE [20] and SOT [21] as well as AHE [5] are opposite to those of CoFeB by using independent SOT and spin pumping measurements. If AUMR originates from the spin accumulation established in the single ferromagnet because of these spin-orbit effects, it is also expected that AUMR reverses in the NiFe and CoFeB. Figure 4(c) and 4(d) present the ΔR and $\Delta R_H$ measurement results for the 10 nm NiFe single layer with $l = 50$ μm and $w = 5$ μm, respectively. Remarkably, the estimated $\Delta R_{max}^{Thermal}$ (23.20 mΩ) is much larger than that of the measured $\Delta R_{max}$ (10.78 mΩ) in the NiFe layer, sharply contrast to the CoFeB layer in which $\Delta R_{max}^{Thermal} < \Delta R_{max}$ (Fig. 3(a) and 3(b)). The fact that $\Delta R_{max}^{Thermal} > \Delta R_{max}$ ($\frac{\Delta R_{max}^{Thermal}}{\Delta R_{max}} > 1$) for NiFe while $\Delta R_{max}^{Thermal} < \Delta R_{max}$ ($\frac{\Delta R^{Thermal}}{\Delta R_{max}} < 1$) for CoFeB single layers has also been confirmed by other different $l/w$ ratios, as summarized in Table 1. These results can only be explained by using a negative and a positive $\Delta R^{AUMR}$ for the NiFe and CoFeB single layer, respectively. Other effects such as non-uniform current distributions may change the thermal contributions by comparing $\Delta R_H^{max}$ with different $l/w$ ratios, but it does not alter the sign of calculated $\Delta R^{AUMR}$, as shown in Table 1. The solid lines in Fig. 4(c) are the corresponding fitting results by considering the thermal effects and AUMR, where the estimated AUMR is about -12.42 mΩ. It should be noted that $\Delta R_H$ shows the same sign for both CoFeB and NiFe (Fig. 3(b) and Fig. 4(d)) and the opposite sign of $\Delta R^{AUMR}$ does indicate that AUMR is reversal in the NiFe and CoFeB, consistent with the mechanism illustrated in Fig. 4(a).

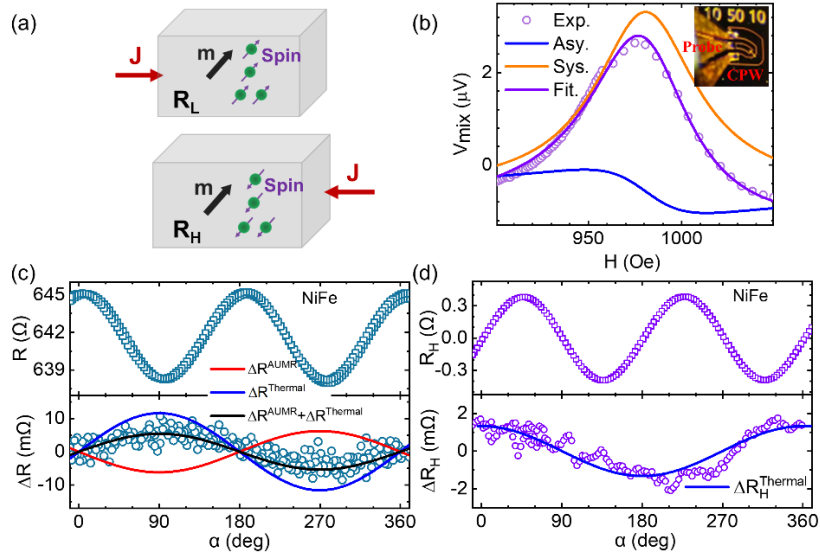

FIG. 4. AUMR mechanisms and AUMR measurement results in NiFe single layers. (a) Illustration of AUMR mechanisms. $R_L$ and $R_H$ represent two different resistance states when the spin polarization generated in the ferromagnet through spin-orbit coupling is parallel and antiparallel to the magnetization, respectively. (b) ST-FMR spectrum of 8 GHz for the 10 nm NiFe single layer. The solid lines are the fitting results by using a symmetric and an antisymmetric Lorentzian curve. The inset shows the optical image of ST-FMR measurement setups. The angle between applied in-plane magnetic field and NiFe strip (10 μm × 50 μm) is 45°. (c), (d) Angle dependence of longitudinal (c) and transverse resistance (d) measurement results for the 10 nm single NiFe layer. The experimental and fitting procedures are exactly the same as those of CoFeB samples.

In a single ferromagnet with bulk inversion symmetry, the spin accumulation generated by spin-orbit coupling should not result in any non-zero bulk spin-orbit effects [38,39]. However, it has been shown that the bulk inversion symmetry can be broken in the as-deposited thin films grown by magnetron sputtering, which is evidenced by observing deterministic spin-orbit switching [44] or pure SOTs [45,46] in the nominally uniform thin films. To verify the possible broken inversion symmetry in the NiFe single layers, the spin-torque ferromagnetic resonance (ST-FMR) [47] is adopted to measure the bulk SOTs. Figure 4(b) presents the ST-FMR spectrum of 8 GHz for the 10 nm NiFe single layer. The dimension of NiFe positioned in the coplanar waveguide (CPW) is 10 μm wide and 50 μm long, and the microwave with the power of 5 dBm is applied through an RF probe, as shown in the inset of Fig. 4(b). Interestingly, even for the NiFe single layer without any adjacent HM layers, the ST-FMR spectrum still shows a sizable resonant peak, which can be fitted by the sum of a symmetric and an antisymmetric Lorentzian curve (Fig. 4(b)). According to the ST-FMR theory, the symmetric and antisymmetric components correspond to the damping- and field-like SOTs [47], respectively. Moreover, the ST-FMR peak is dominated by the symmetric component, indicating that the strong damping-like SOT, which can only arise from self-generated spin accumulation, does exist in the NiFe single layer. Therefore, the ST-FMR results not only provide clear evidences for the existence of broken inversion symmetry, but also demonstrate the appearance of spin accumulation in the single ferromagnetic layer. However, for the CoFeB single layer, we did not observe the ST-FMR peak, probably because the AMR of CoFeB ($\approx$ 0.12% from Fig. 3(a)) is much weaker than that of NiFe ($\approx$ 1.08% from Fig. 4(c)) by considering that the strength of ST-FMR signals is determined by the magnitude of AMR in the single ferromagnetic layer.

Recently, it has been reported the non-uniform spatial distribution of elements along the thickness direction could induce inversion symmetry breaking and create non-zero bulk Dzyaloshinskii–Moriya interaction in a single amorphous layer [48]. Therefore, to directly demonstrate the possible inversion symmetry breaking in the CoFeB layer, the structure and element distribution of the 10 nm CoFeB single layer are investigated by using HRTEM and EDS. As shown in Fig. 5(a), no crystalline lattice can be found in the CoFeB layer as expected since the sputtered alloys are usually amorphous. The corresponding element mapping results presented in Fig. 5(b) show that the Co, Fe, and B elements distribute in the entire CoFeB region and the B element has diffused into the adjacent $SiO_2$ layers. The O mapping reflects O distributions in the $SiO_2$ substrate and capping layer. To precisely determine the Co : Fe atomic ratios, EDS is performed at the selected sites along the thickness ($z$) direction. We choose two positions, $P_1$ and $P_2$ arrows as marked in Fig. 5(c), and select five sites along the $z$ direction at each position to measure the atomic ratios. Figure 5(d) presents the detected Co : Fe atomic ratio along the thickness direction. One can see that the Co : Fe ratio decreases monotonously from the site close to the substrate to the $SiO_2$ capping layer. The monotonous decrease of Co : Fe ratio may be the origin of inversion symmetry breaking in the CoFeB layer, as demonstrated in other amorphous alloys [48]. Note that the boron content cannot be detected accurately by EDS, and the detected Co : Fe atomic ratio varies around 1 : 1, which is close to that in the $Co_4Fe_4B_2$ sputtering target. On the other hand, if the broken bulk inversion symmetry is due to the non-uniform element distribution, it can be modulated by the annealing processes, and thus the AUMR should also be altered. As shown in Table 1, for the 5 nm CoFeB layer annealed at 200 °C for 10 minutes, the thermal effects due to Joule heating ($\Delta R_H^{max}$) only change slightly from 17.88 mΩ to 14.77 mΩ, while $\Delta R_{max}$ drops dramatically from 108.74 mΩ to 78.57 mΩ due to the reduction of $\Delta R^{AUMR}$. These results further confirm the existence of broken bulk inversion symmetry and resultant AUMR in the single ferromagnetic layers. Moreover, it is expected that the magnitude of $\Delta R^{AUMR}$ depends on the degree of broken inversion symmetry and spin-orbit coupling strength of the ferromagnetic materials.

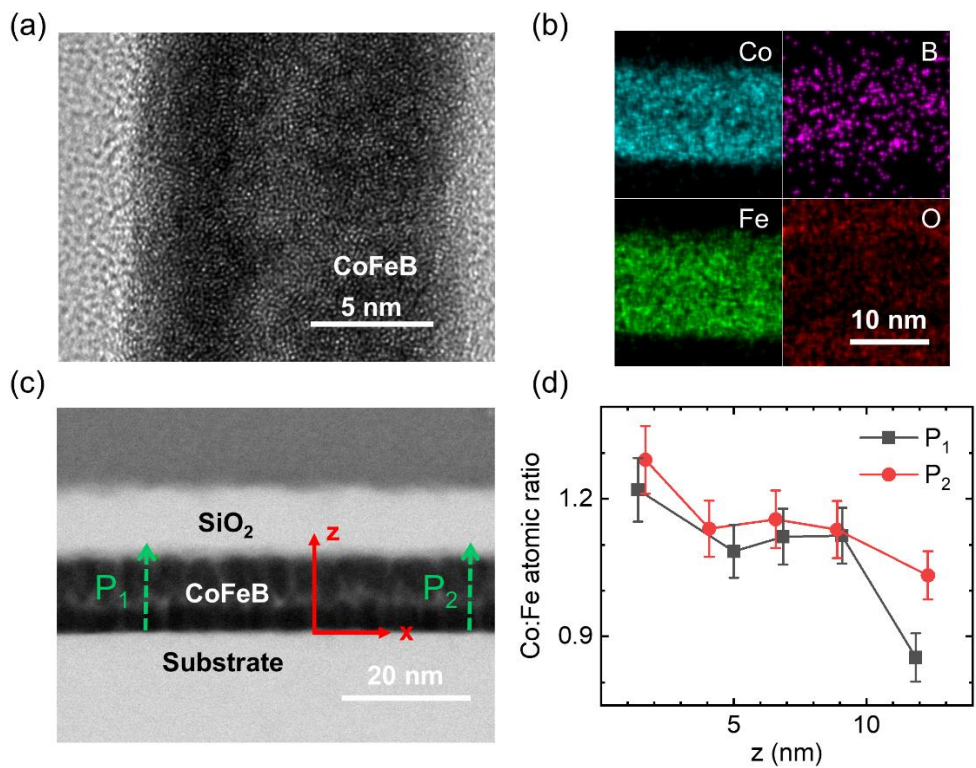

FIG. 5. HRTEM and element mapping results of the 10 nm CoFeB single layer. (a) HRTEM images and (b) Co, Fe, B, and O elemental maps of the CoFeB single layer. (c) The STEM image for performing EDS measurements. $P_1$ and $P_2$ arrows indicate two lines for acquiring EDS signals along the thickness direction. (d) The measured Co : Fe atomic ratio as a function of $z$ along the P1 and P2 arrows.

In application, the observed large AUMR can be facilely used to detect in-plane magnetization switching with no need of complex MTJ structures and lock-in techniques, especially for the materials that are not suitable for fabricating MTJs. Figure 6 shows the current driven in-plane magnetization switching detected by measuring AUMR in Pt (6)/CoFeB (3) structures. The two stable magnetization states at zero field are first confirmed by measuring ΔR as a function of in-plane external magnetic field, in which ΔR signals correspond to the two in-plane magnetization states separate clearly as shown in Fig. 6(a). Figure 6(b) presents ΔR variation after applying the switching current pulses. It can be seen that, even for the gradual in-plane magnetization switching, each step due to the formation of multi-domains can also be distinguished by using the common dc measurements. Since the AUMR of CoFeB does not depend on the neighboring SOT layers, the in-plane magnetization detection techniques can be conveniently expanded to explore the SOT switching in other SOT material-involved systems with the same sensitivity [24–27].

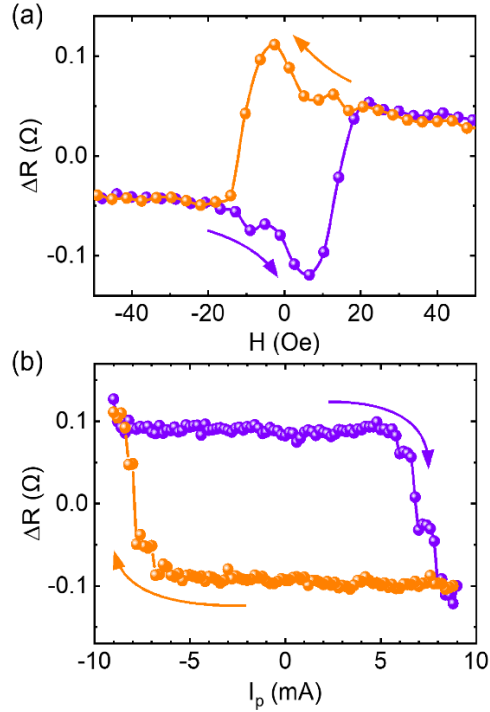

FIG. 6. In-plane magnetization switching detected by AUMR. (a) ΔR as a function of applied in-plane magnetic field for the Pt (6)/CoFeB (3) devices $l = 50$ μm and $w = 2$ μm. (b) Current-driven multidomain switching of in-plane magnetization measured by recording AUMR as a function of driving current pulse, $I_p$. The arrows indicate the field or current sweep directions.

## IV. Conclusions

The counterpart of GMR in a single ferromagnet due to spin accumulation generated by an applied in-plane current has been clearly demonstrated in experiments. By directly calibrating temperature increase due to Joule heating and corresponding thermal voltage generated by ANE and SSE, we have quantitatively extracted AUMR of a single ferromagnet. Our results show that the AUMR signals, after exclusion of thermal contribution, are reversal for the CoFeB and NiFe single layers, where the signs of SOT in both materials are opposite, confirming that the AUMR originates from the same fundamental mechanisms as other spin-orbit effects such as SHEs and SOTs. The ST-FMR and HRTEM techniques as well as annealing processes are further performed to demonstrate that the observed non-zero bulk spin-orbit effects arise from broken bulk inversion symmetry induced during sample deposition. From the viewpoint of practical applications, the large AUMR provides a reliable and simple electrical strategy for reading magnetization information in the two-terminal magnetic devices to avoid the technically challenging TMR multilayers.

*Note added.* We noticed that the so-called anomalous-Hall unidirectional magnetoresistance (AH-UMR) is predicted theoretically [49], which agrees with our experimental results that such UMR originates from spin-orbit coupling of the single ferromagnet itself.

**Acknowledgments**

This work is supported by the National Key R&D Program of China (Grant No. 2019YFB2005800 and 2018YFA0701500), the National Natural Science Foundation of China (Grant No. 61974160, 61821091, and 61888102), and the Strategic Priority Research Program of the Chinese Academy of Sciences (Grant No. XDB44000000). We thank Steven S.-L. Zhang for helpful discussions.